\documentclass[useAMS,usenatbib]{mn2e}

\usepackage{epsfig}
\usepackage{amsmath}
\usepackage{amssymb}
\usepackage{natbib}
\usepackage{threeparttable} 
\usepackage{multirow}

\usepackage{epstopdf}

\newcommand{\chan}{\textit{Chandra}}
\newcommand{\swift}{\textit{Swift}}
\newcommand{\rxte}{\textit{RXTE}}

\newcommand{\inte}{\textit{Integral}}
\newcommand{\maxi}{\textit{MAXI}}
\newcommand{\exosat}{\textit{EXOSAT}}
\newcommand{\rosat}{\textit{ROSAT}}

\newcommand{\hakucho}{\textit{Hakucho}}

\newcommand{\Msun}{\mathrm{M}_{\odot}}
\newcommand{\lum}{\mathrm{erg~s}^{-1}}
\newcommand{\flux}{\mathrm{erg~cm}^{-2}~\mathrm{s}^{-1}}
\newcommand{\cnts}{\mathrm{counts~s}^{-1}}
\newcommand{\mdot}{\mathrm{M_{\odot}~yr}^{-1}}
\newcommand{\nh}{\mathrm{cm}^{-2}}

\newcommand{\exo}{EXO 1745--248}
\newcommand{\igr}{IGR J17480--2446}

\newcommand{\sax}{SAX J1808.4--3658}

\newcommand{\xte}{XTE J1701--462}
\newcommand{\qpexo}{EXO~0748--676}
\newcommand{\ks}{KS~1731--260}
\newcommand{\mxb}{MXB~1659--29}

\def \mnras {MNRAS}
\def \apj {ApJ}
\def \apjs {ApJS}
\def \apjl {ApJL}
\def \aap {A\&A}

\def \araa {ARAA}

\def \pasj {PASJ}

\def \iaucirc {IAU Circ.}

\title[EXO 1745--248 in quiescence]{Strong X-ray variability in the quiescent state of the neutron star low-mass X-ray binary EXO 1745--248}
\author[N. Degenaar \& R. Wijnands]
{N. Degenaar$^{1,2}$\thanks{e-mail: degenaar@umich.edu}\thanks{Hubble fellow} \&
R. Wijnands$^1$\\
$^1$Astronomical Institute "Anton Pannekoek", 
University of Amsterdam, 
Postbus 94249, 1090 GE Amsterdam, The Netherlands\\
$^2$University of Michigan, 
Department of Astronomy, 
500 Church St, Ann Arbor, MI 48109-1042, USA
}

\begin{document}

\date{Accepted 2012 January 25. Received 2012 January 25; in original form 2011 August 4}

\pagerange{\pageref{firstpage}--\pageref{lastpage}} \pubyear{0000}

\maketitle

\label{firstpage}

\begin{abstract} 
The transient neutron star low-mass X-ray binary \exo, located in the globular cluster Terzan 5, was detected during its quiescent state with \chan\ in 2003. The source displayed a 0.5--10 keV quiescent X-ray luminosity of $L_{q}\sim10^{33}~(D/5.5~\mathrm{kpc})^2~\lum$, which was completely dominated by hard non-thermal emission. This is at odds with other non-pulsating neutron stars that typically show detectable soft thermal emission at such quiescent luminosities. Here we use three additional \chan\ observations, performed in 2009 and 2011, to further study the quiescent properties of \exo. We find that the powerlaw intensity varies considerably up to a factor of $\sim 3$ within hours and by about one order of magnitude between the different epochs. We discuss the implications of the observed change in quiescent flux for the interpretation of the hard powerlaw emission. We constrain the neutron star surface temperature as seen by a distant observer to $kT^{\infty}\lesssim42$~eV and the thermal bolometric luminosity to $L_{\mathrm{q,bol}}\lesssim 7 \times10^{31}~(D/5.5~\mathrm{kpc})^2~\lum$. This confirms that \exo\ harbours a relatively cold neutron star and suggests that, for example, enhanced cooling mechanisms are operating in the stellar core or that the binary on average resides in quiescence for hundreds of years.
\end{abstract}


\section{Introduction}
Low-mass X-ray binaries (LMXBs) are binary star systems that are composed of a compact primary (a neutron star or a black hole) that accretes the outer gaseous layers of a (sub-) solar-mass companion star. Neutron star primaries can reveal themselves by showing coherent X-ray pulsations when the accretion flow is funnelled onto the magnetic poles of the neutron star, or by displaying type-I X-ray bursts caused by unstable thermonuclear burning of the accreted matter on the surface of the neutron star. 

Many neutron star LMXBs are transient and alternate active accretion outbursts with long episodes of quiescence. Whereas outbursts typically last a only few weeks and generate 0.5--10 keV X-ray luminosities of $L_X\sim10^{36-38}~\lum$ \citep[e.g.,][]{chen97}, the quiescent phase may extend to many decades and is characterised by a much lower quiescent luminosity of $L_q\sim10^{31-33}~\lum$ \citep[0.5--10 keV;][]{heinke2009}. The cause of the quiescent X-ray emission has been subject of debate \citep[e.g.,][]{campana2003}. Spectrally, one can distinguish a soft thermal component at energies below $\sim2$~keV, and a hard non-thermal emission tail that dominates the quiescent X-ray spectrum above $\sim2-3$ keV. 

The soft quiescent spectral component is generally interpreted as heat being radiated from the surface of the cooling neutron star \citep{brown1998,colpi2001}. As such, the quiescent thermal emission provides a measure of the neutron star's interior temperature. However, low-level accretion onto the surface of the neutron star might also produce a thermal X-ray spectrum \citep[][]{zampieri1995}. 

The hard non-thermal emission is usually modelled as a simple power law with a photon index of $\Gamma \sim1-2$, and its fractional contribution to the total 0.5--10 keV flux may vary anywhere from $0-100\%$ \citep[][]{jonker2004}. It has been attributed e.g., to shock emission generated in the interaction of a residual accretion flow with the magnetic field of the neutron star, or the re-activation of a radio pulsar in quiescence \citep{campana1998}. It remains an unanswered question if, and how, the two spectral components are related.

\begin{table*}
\caption{Historic activity from Terzan 5.}
\begin{threeparttable}
\begin{tabular}{l l l l l l l}
\hline \hline
Year & Satellite(s) & $L_X$ & $L_{X,\mathrm{peak}}$ & $t_{\mathrm{ob}}$ & Comments & Ref. \\
 & & ($\lum$) & ($\lum$) & (days) &  & \\
\hline
1980 & \hakucho\ & $\lesssim 3 \times 10^{36}$ & - & $> 16$ & Type-I X-ray bursts & 1,2 \\
1984 & \exosat\ & $\sim 3 \times 10^{37}$ & - & - & & 3 \\
1990 & \rosat\ (PSPC) & $\sim 3\times10^{35}$ & - & - & & 4 \\
1991 & \rosat\ (HRI) & $\sim 3 \times 10^{34}$ & - & - & & 5 \\
2000 & \rxte, \chan\ & $\sim 2 \times 10^{37}$ & $\sim 6 \times 10^{37}$ & $\sim100$  & Type-I X-ray bursts, dipping & 6,7,8 \\
2002 & \rxte\ & $\sim 2\times 10^{37}$ & $\sim 4 \times 10^{37}$ & $\sim30$ &  & 8,9 \\  
2010 & \inte, \rxte, \swift, \chan & $\sim 2 \times 10^{37}$ & $\sim 7 \times 10^{37}$ & $\sim80$ & Type-I X-ray bursts, 11-Hz pulsations & 10-15\\
2011 & \maxi, \rxte, \swift, \chan\ & $\sim 3 \times 10^{36}$ & $\sim 9 \times 10^{36}$ & $\sim20$ & Superburst & 16-19\\
\hline
\end{tabular}
\label{tab:history}
\begin{tablenotes}
\item[]Note. -- $L_X$ represents the (average) 0.5--10 keV luminosity and $t_{\mathrm{ob}}$ reflects any available constraints on the outburst duration. We calculated $L_X$ from fluxes reported in literature or using publicly available \rxte/ASM or \maxi\ data. For the conversion to the 0.5--10 keV energy band we use \textsc{PIMMS} (v. 4.4) by assuming a fiducial spectrum with $N_H= 1 \times10^{22}~\nh$ and $\Gamma=1.5$, and we adopt a distance of $D=5.5$~kpc. When possible we also give the 0.5--10 keV peak luminosity, $L_{X,\mathrm{peak}}$. References: 1=\citet{makishima1981}, 2=\citet{inoue1984}, 3=\citet{warwick1988}, 4=\citet{verbunt1995_rosat}, 5=\citet{johnston1995}, 6=\citet{markwardt2000}, 7=\citet{heinke2003}, 8=Altamirano et al. in preparation, 9=\citet{wijnands2002_terzan5}, 10=\citet{bordas2010}, 11=\citet{papitto2010}, 12=\citet{miller2011}, 13=\citet{motta2011}, 14=\citet{linares2011}, 15=\citet{deeg_wijn2011_2}, 16=\citet{altamirano2011a}, 17=\citet{altamirano2011b}, 18=\citet{mihara2011}, 19=\citet{pooley2011}.
\end{tablenotes}
\end{threeparttable}
\end{table*}

\subsection{\exo\ in Terzan 5}
Terzan 5 is a globular cluster located in the bulge of our Galaxy at an estimated distance of $D=5.5$~kpc \citep{ortolani2007}.\footnote{Throughout this work we assume a distance of $D=5.5$~kpc for Terzan 5 and \exo.} This cluster has a particularly high stellar density, which is thought to cause a high rate of dynamical star encounters and the formation of compact binary systems \citep[e.g.,][]{pooley2003}. Apart from a large number of millisecond radio pulsars \citep[][]{ransom2005} and faint X-ray point sources that may be quiescent LMXBs or cataclysmic variables \citep[][]{heinke2006_terzan5}, Terzan 5 also contains two confirmed transiently accreting neutron stars. 

\begin{figure*}
 \begin{center}
\includegraphics[width=15.0cm]{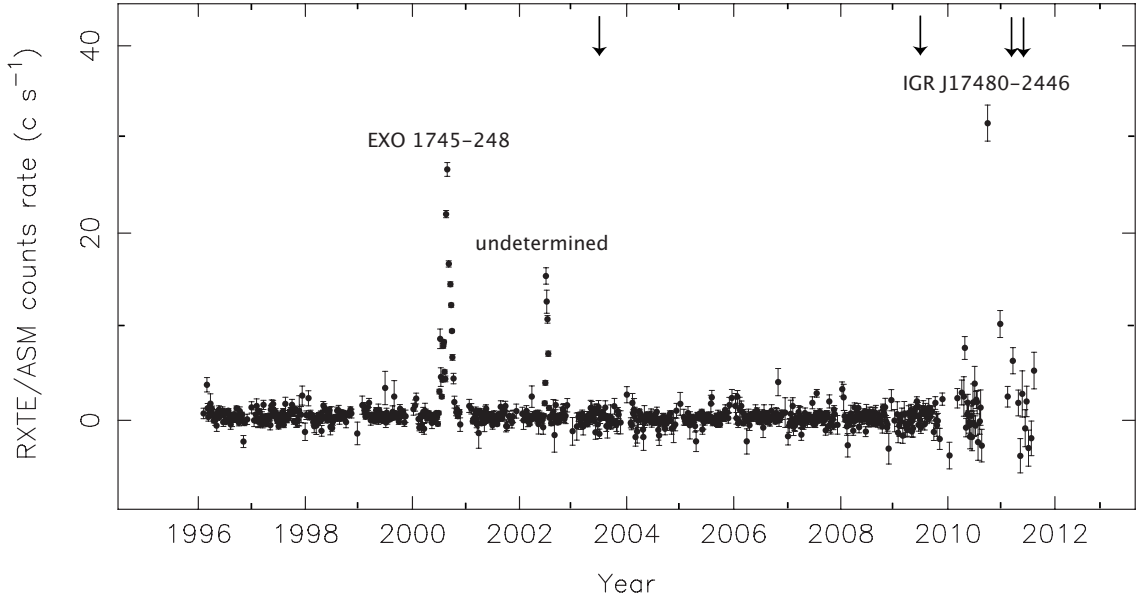}	
    \end{center}
\caption[]{{\rxte/ASM 5-d averaged lightcurve of Terzan 5, illustrating three different outbursts that were recorded in the past 15 years (1.5--12 keV). Arrows indicate \chan\ observations that were performed when no bright X-ray transients were active (Table~\ref{tab:obs}). We note that the most recent outburst that was detected in 2011 October--November (\exo) was not covered by the \rxte/ASM.}}
 \label{fig:asm}
\end{figure*}

Already in 1980, a number of type-I X-ray bursts was detected from the direction of the cluster, which testified the presence of an active neutron star LMXB \citep[][]{makishima1981}. Subsequent X-ray activity was observed from Terzan 5 in 1984, 1990, 1991, 2000, 2002, 2010 and most recently in 2011 (Table~\ref{tab:history} and references therein). While the detections in 1984, 1990 and 1991 concern single observations and hence only snapshots of the activity, the most recent four outbursts have been covered by several X-ray instruments, e.g., by the ASM onboard \rxte\ (Fig.~\ref{fig:asm}). 

Multiple type-I X-ray bursts were detected during the 2000 and 2010 outbursts, while a superburst (i.e., a very energetic thermonuclear burst that lasts for several hours) was observed in 2011. This thus evidences active accretion onto a neutron star in 2000, 2010 and 2011 \citep[][]{galloway2008,linares2011,motta2011,altamirano2011b,mihara2011}. High-spatial resolution \chan\ observations have revealed that these three outbursts involved two different transient neutron star LMXBs \citep[cf.][]{heinke2003,pooley2010,pooley2011}. 

Historically, the source that was active in 2000 and 2011 is referred to as \exo, whereas the recently discovered 2010 transient has been dubbed \igr.\footnote{In the study of \citet{heinke2006_terzan5}, \exo\ is denoted as CX3 and \igr\ as CX25.} In addition to exhibiting type-I X-ray bursts, the latter also displays 11-Hz X-ray pulsations \citep{strohmayer2010,papitto2010}. 
The two transient neutron star LMXBs have a spatial separation of merely $\sim5''$ (see Fig.~\ref{fig:ds9}). Hence, apart from \chan, none of the present and past X-ray missions provides sufficient angular resolution to spatially resolve them. It is thus unclear which of the two transient neutron star LMXBs was causing the outbursts that occurred in 1980, 1984, 1990, 1991 and 2002. Moreover, 50 distinct low-luminosity X-ray point sources have been identified within the half-mass radius of Terzan 5 \citep[$0.83'$;][]{heinke2006_terzan5}, so the possibility of additional active transient sources causing these historic outbursts cannot be ruled out.

Terzan 5 has been observed with \chan\ on several occasions when no bright transients were active (Table~\ref{tab:obs}). These observations can be addressed to investigate the two neutron star LMXBs during quiescence. The quiescent properties of the 2010 transient, \igr, are the subject of a recent ongoing study \citep{deeg_wijn2011_2,deeg_wijn2011,degenaar2011_terzan5_3}. The quiescent emission spectrum of the 2000/2011 transient, \exo, was studied by \citet{wijnands2005} using a \chan\ observation obtained in 2003. At that time the source displayed a 0.5--10 keV luminosity of $L_q\sim10^{33}~\lum$, which is not unusual for neutron star LMXBs in quiescence \citep[e.g.,][]{heinke2009}. Unlike other non-pulsating neutron star LMXBs, however, the quiescent X-ray spectrum of \exo\ was found to be completely dominated by a hard emission component, with no indications of thermal radiation \citep{wijnands2005}. 
Although a non-thermal component is often seen for neutron star LMXBs in quiescence \citep[e.g.,][]{jonker2004}, typically thermal emission can be detected as well. 

The lack of detectable quiescent thermal emission places strong constraints on the properties of the neutron star core \citep{jonker2004_saxj1810,jonker07,tomsick2005,wijnands2005,heinke2009}. These neutron stars must be cold, suggesting that their interior may be efficiently cooled through neutrino emissions (see also Sec.~\ref{sec:thermal}). Such neutron stars are expected to be relatively massive, because larger central densities are thought to lead to a higher rate of neutrino cooling \citep[][]{lattimer2004,yakovlev2004}. 
In this work, we discuss three additional \chan\ observations of Terzan 5, carried out in 2009 and 2011, to further investigate the quiescent properties of \exo.

\section{Chandra observations of Terzan 5}
We use four \chan\ observations of Terzan 5, spread between 2003 and 2011, to study any possible spectral and temporal variability in the quiescent emission of \exo. Details of the individual observations can be found in Table~\ref{tab:obs} and references therein. All four were carried out in the faint data mode, with the nominal frame time of 3.2~s and the target positioned on the S3 chip. Data reduction was carried out using the \textsc{ciao} tools (v. 4.3) and following standard \chan\ analysis threads.\footnote{http://cxc.harvard.edu/ciao/threads/index.html.} The 2003 and 2009 data were reprocessed using the task ACIS$\_$PROCESS$\_$EVENTS to benefit from the most recent calibration. 
Episodes of high background were removed from the 2003 data, which resulted in a net exposure time of 31.2~ks. No background flares were present during the 2009 and 2011 observations, so the full exposure time was used in the analysis. 

\begin{figure*}
 \begin{center}
\includegraphics[width=8.0cm]{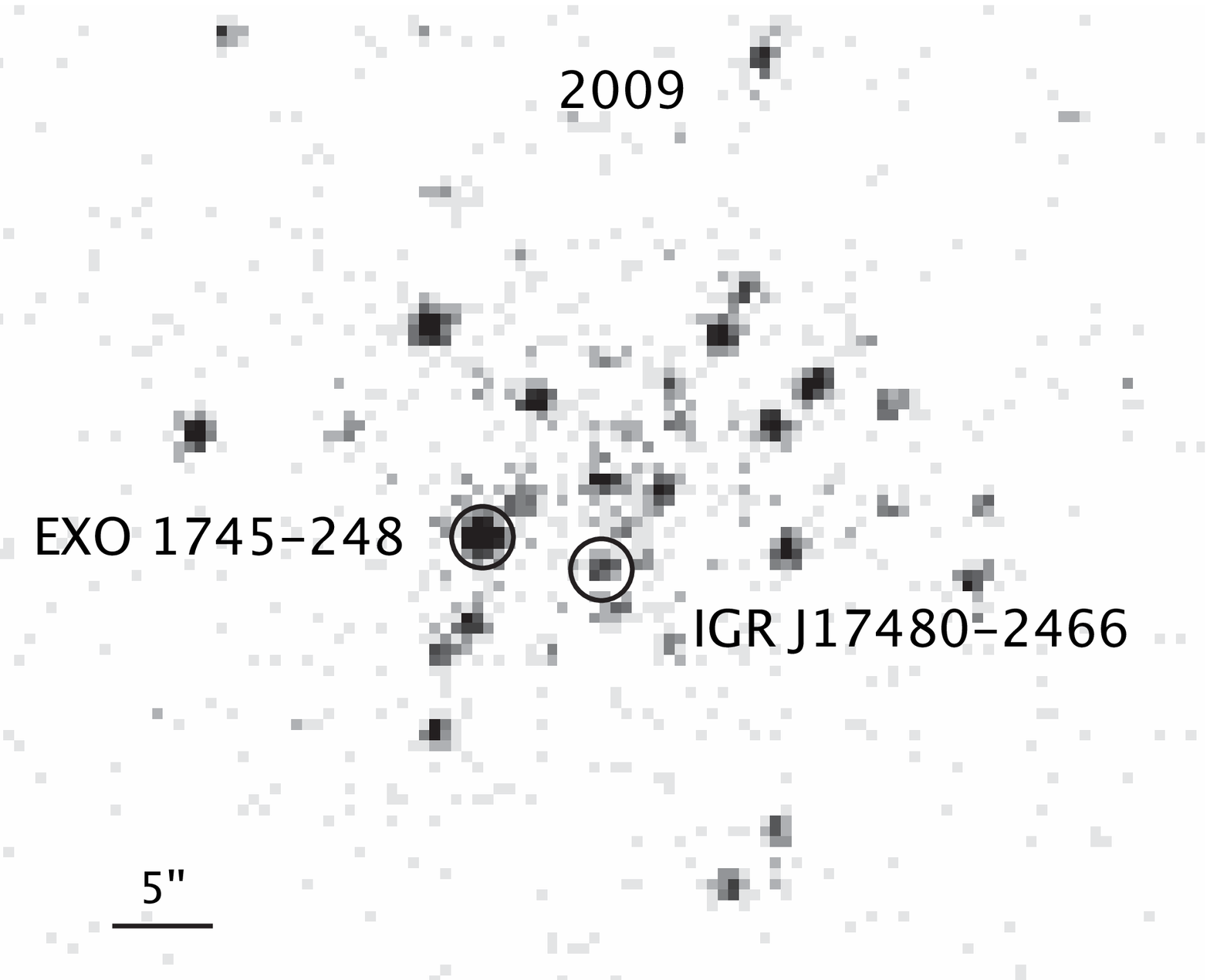}\hspace{0.5cm}
 \includegraphics[width=8.0cm]{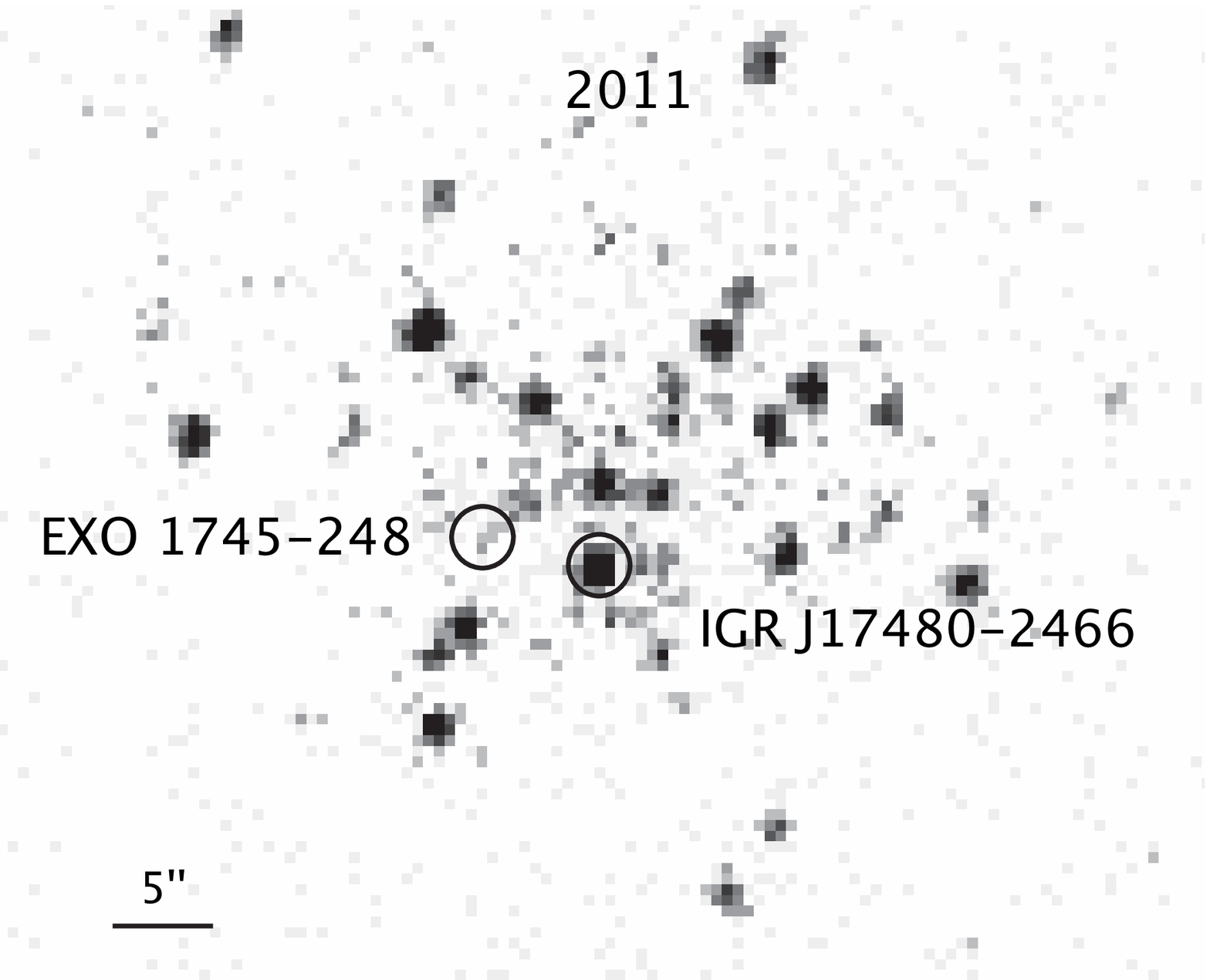}
    \end{center}
\caption[]{{\chan/ACIS images obtained in 2009 and 2011 (0.5--5 keV). For the 2011 image, we summed the two individual exposures of February and April (Table~\ref{tab:obs}). The locations of the two known transient neutron star LMXBs are indicated by circles of $1.5''$ radii.
}}
 \label{fig:ds9}
\end{figure*}

\exo\ is clearly detected in the dense core of the cluster in 2003 and 2009 (Fig.~\ref{fig:ds9}, left). We used a circular region with a $1.5''$ radius to extract source events, and one with a radius of $40''$ positioned on a source-free part of the CCD $\sim1.4'$ west of the cluster core as a background reference. We extracted count rates and lightcurves using the tool DMEXTRACT, while the meta-task SPECEXTRACT was used to obtain spectra and to generate the ancillary response files (arf) and redistribution matrix files (rmf). The spectral data was grouped into bins with a minimum of 15 photons using the tool GRPPHA and fitted using the software package \textsc{XSpec} (v. 12.7). To take into account the interstellar neutral hydrogen absorption along the line of sight, we include the PHABS model in all our spectral fits using the default \textsc{XSpec} abundances and cross-sections. Throughout this work we assume a distance of $D=5.5$~kpc towards \exo\ when converting X-ray fluxes into luminosities. All quoted errors correspond to $90\%$ confidence levels.

Whereas \exo\ was one of the brightest sources in Terzan 5 during the 2003 and 2009 observations, the source had considerably faded in 2011 (Fig.~\ref{fig:ds9}). Although the source is not clearly visible in the 2011 data sets, there appears to be an excess of photons present at the source position. To verify this, we employed the wavelet detection algorithm PWDETECT \citep{damiani1997_1,damiani1997_2}, which has been found to be particularly effective in detecting faint sources located close to brighter objects and hence it is a very useful tool for globular clusters \citep[e.g.,][]{heinke2006_terzan5}. We performed standard PWDETECT runs on the ACIS-S3 chip with wavelet sizes varying from $0.5''-2.0''$.\footnote{See http://www.astropa.unipa.it/progetti$\_$ricerca/PWDetect.} This exercise suggests that \exo\ was weakly detected during both 2011 observations (see Sect.~\ref{subsec:2011}). We extracted count rates and spectra using the tools DMEXTRACT and SPECEXTRACT, respectively. We combined the 2011 spectral data using the \textsc{ciao}-task COMBINE$\_$SPECTRA to improve the statistics.

\begin{table}
\caption{\chan\ observations of \exo\ in quiescence.}
\begin{threeparttable}
\begin{tabular}{l l l l l}
\hline \hline
Date & Obs ID & $t_{\mathrm{exp}}$ & Net count rate & Ref. \\
(yyyy-mm-dd) & & (ks) & ($\cnts$) & \\
\hline
2003-07-13/14 & 3798 & 39.5 & $7.1\times10^{-3}$ & 1,2,3 \\ 
2009-07-14/15 & 10059 & 36.4 & $1.1\times10^{-2}$ & 3 \\ 
2011-02-17 & 13225 & 29.7 & $4.0\times10^{-4}$ & 4 \\
2011-04-29/30 & 13252 & 39.5 & $3.8\times10^{-4}$ & 5 \\ 
\hline
\end{tabular}
\label{tab:obs}
\begin{tablenotes}
\item[]Note. -- After correcting for background flares, the exposure time of the 2003 observation is $t_{\mathrm{exp}}=31.2$~ks. The quoted net source count rates are for the 0.5--10 keV band. Observation details can be found in the references: 1=\citet{heinke2006_terzan5}, 2=\citet{wijnands2005}, 3=\citet{deeg_wijn2011}, 4=\citet{deeg_wijn2011_2}, 5=\citet{degenaar2011_terzan5_3}.
\end{tablenotes}
\end{threeparttable}
\end{table}

\section{\exo\ in quiescence}
\subsection{The 2003/2009 data: spectra and variability}\label{subsec:spec}

The 2003 \chan\ observation of Terzan 5 was used by \citet{wijnands2005} to study the quiescent properties of \exo. At that time the source was detected at a mean count rate of $(0.71\pm0.05)\times10^{-2}~\cnts$ and displaying significant variability in the light curve (0.5--10 keV). The quiescent spectrum was found to be hard and absent of a thermal emission component. The data could be best described by a single absorbed powerlaw model with a hydrogen column density $N_H=(1.4\pm0.5)\times10^{22}~\nh$ and a spectral index of $\Gamma=1.8\pm0.5$. 
\citet{wijnands2005} concluded that any possible thermal emission component contributed at most $\sim10\%$ to the unabsorbed 0.5--10 keV flux of $F_X=2.2^{+0.7}_{-0.3}\times10^{-13}~\flux$. An upper limit on the neutron star temperature as seen by an observer at infinity of $kT^{\infty}\lesssim80$~eV was inferred. We reprocessed and refitted the 2003 data in this work and obtained similar results as found previously (see Table~\ref{tab:spec}).

\begin{figure}
 \begin{center}
 \includegraphics[width=8.0cm]{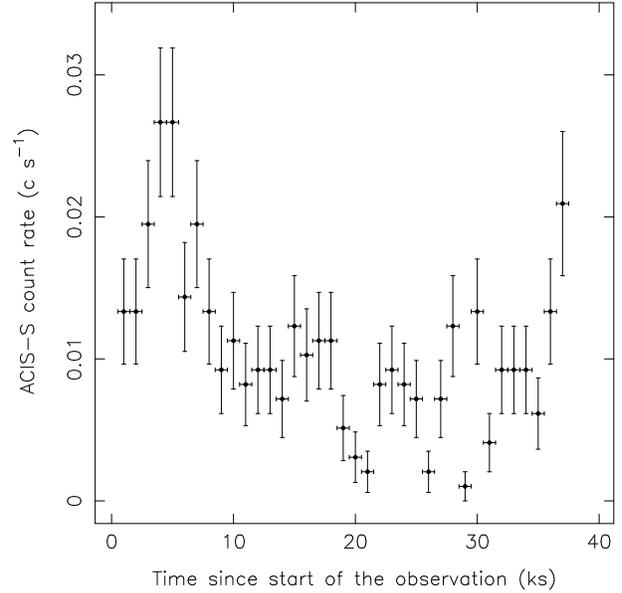}
    \end{center}
\caption[]{{X-ray count rate lightcurve of \exo\ obtained from the 2009 \chan\ observation of Terzan 5, covering the energy range of 0.5--10 keV and using a bin time of 1000~s. }}
 \label{fig:lc}
\end{figure}

During the 2009 observation \exo\ was detected at an average count rate of $(1.09\pm0.06)\times10^{-2}~\cnts$ (0.5--10 keV). The count rate lightcurve, shown in Fig.~\ref{fig:lc}, reveals considerable variation by a factor of $\sim3$ on a time scale of hours. We fitted the 2009 spectral data using a variety of models. Similar to the results for the 2003 observation, we find that a thermal emission model does not provide a good description of the data. We first tried a neutron star atmosphere model, which is often used to describe the X-ray spectra of quiescent neutron star LMXBs. Using the model NSATMOS \citep[][]{heinke2006} we explored fits by either fixing several parameters or leaving these free to float. All trials resulted in a reduced chi-squared value $\chi_{\nu}^2>6$, hence no acceptable fit could be obtained. 

A simple blackbody can describe the data, but returns unrealistic model parameters: $N_H<0.3\times10^{22}~\nh$, a temperature $kT=1.4\pm0.2$~keV and an emitting radius $R=0.04^{+0.6}_{-0.04}$~km (for $D=5.5$~kpc), yielding $\chi_{\nu}^2=1.1$ for 21 degrees of freedom (dof). The obtained hydrogen column density is well below that inferred for the 2000 outburst and 2003 quiescent data of \exo\ \citep[$N_H \sim 1.5 \times10^{22}~\nh$;][]{heinke2003,wijnands2005}, as well as the average value found for the X-ray sources in Terzan 5 \citep[$N_H \sim 1.9 \times10^{22}~\nh$;][]{heinke2006}. Furthermore, the temperature and emitting radius would be highly unusual for a (non-pulsating) thermally-emitting neutron star in quiescence \citep[e.g.,][]{rutledge1999}. We thus conclude that this model does not correctly describe the data.

Fitting the 2009 spectral data with a simple absorbed powerlaw yields $N_H=(1.1\pm0.3)\times10^{22}~\nh$ and $\Gamma=1.3\pm0.3$ ($\chi_{\nu}^2=1.3$ for 22 dof). The resulting  0.5--10 keV unabsorbed flux is $F_X=(2.7\pm0.2) \times10^{-13}~\flux$ (see Table~\ref{tab:spec}). To constrain any thermal emission component, we add an NSATMOS model with a canonical neutron star mass and radius of $M_{NS}=1.4~\Msun$ and $R_{NS}=10$~km, a source distance of $D=5.5$~kpc and a normalisation of unity. The only free fit parameter for this model component is the neutron star temperature. Addition of this model component does not improve the fit ($\chi_{\nu}^2=1.3$ for 21 dof). Since the spectral data do not require the inclusion of a thermal component, we consider the obtained neutron star temperature and thermal flux as upper limits to their true values. As such we find that the thermal component contributes $\lesssim42\%$ to the total 0.5--10 keV unabsorbed flux and we obtain a neutron star temperature of $kT^{\infty}\lesssim85$~eV. By extrapolating the NSATMOS model fit to an energy range of 0.01--100 keV, we estimate a thermal bolometric luminosity of $L^{\mathrm{th}}_{\mathrm{q,bol}}\lesssim 1 \times10^{33}~\lum$. This is comparable to the results for the 2003 observation (see Table~\ref{tab:spec}). 

\begin{figure}
 \begin{center}
\includegraphics[width=8.0cm]{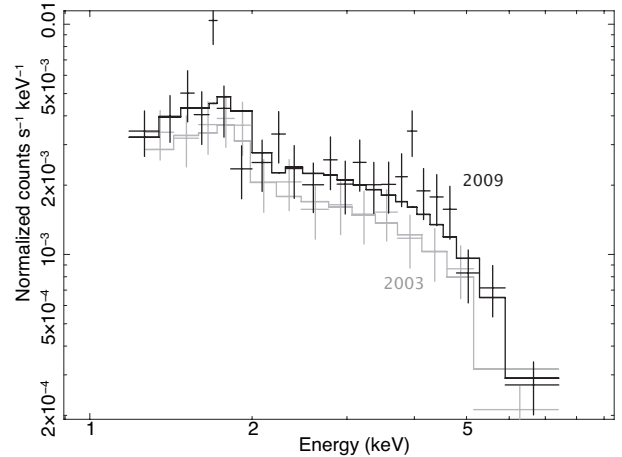}
    \end{center}
\caption[]{{\chan/ACIS spectra of \exo\ as observed in 2003 (bottom, grey) and 2009 (top, black). The solid lines represent the best-fit results for an absorbed powerlaw model with the hydrogen column density and spectral index fixed between the two epochs, while the powerlaw normalisation is variable.}}
 \label{fig:spec}
\end{figure}

The 0.5--10 keV unabsorbed flux inferred from the 2009 spectral data is higher than observed in 2003. This suggests that the source intensity may be variable on a timescale of years, although the two measurements are consistent within the $90\%$ confidence errors (see Table~\ref{tab:spec}). To further investigate whether there are intensity or spectral variations between 2003 and 2009, we fitted the two data sets simultaneously to an absorbed powerlaw model. When all spectral parameters are forced to be the same (i.e., assuming there are no spectral differences between the two data sets), we obtain $N_H=(1.3\pm0.3)\times10^{22}~\nh$ and $\Gamma=1.6\pm0.3$ for $\chi_{\nu}^2=1.3$ (38 dof). Enabling the powerlaw normalisation to vary between the two epochs (i.e., allowing only the intensity to change) provides a better fit that yields $N_H=(1.2\pm0.3)\times10^{22}~\nh$ and $\Gamma=1.5\pm0.3$ for $\chi_{\nu}^2=1.0$ (36 dof). This model fit is shown in Fig.~\ref{fig:spec}. An f-tests suggests that there is a $1\%$ probability of achieving this level of improvement by chance. The inferred 0.5--10 keV unabsorbed fluxes are $F_X=(2.0\pm0.2) \times10^{-13}$ and $(2.7\pm0.2) \times10^{-13}~\flux$ for the 2003 and 2009 data, respectively. This suggests that the source intensity changed by $\sim30\%$ between the two epochs.

We next allowed the spectral shape to vary between 2003 and 2009. The 2000 outburst lightcurve of \exo\ displayed X-ray dips \citep[][]{markwardt2000_2}. Such behavior has been seen in some LMXBs and is thought to result from obscuration of the central X-ray source, e.g., by the outer edge of the accretion disc \citep[e.g.,][]{boirin2005}. As a result, the absorption column density along the line of sight may be variable. We therefore fitted the quiescent data of \exo\ with the powerlaw spectral index fixed between the two epochs, while the hydrogen column density was allowed to vary. The results from this fit suggest that the X-ray spectrum was possibly more absorbed in 2009, although the fit-values obtained for the two epochs are consistent within the errors (see Table~\ref{tab:spec}). Keeping the hydrogen column density fixed and instead allowing the powerlaw index to vary yields a harder spectral index for the 2009 data, although within the errors the value is consistent with the 2003 result (Table~\ref{tab:spec}).

\begin{figure}
 \begin{center}
 \includegraphics[width=8.0cm]{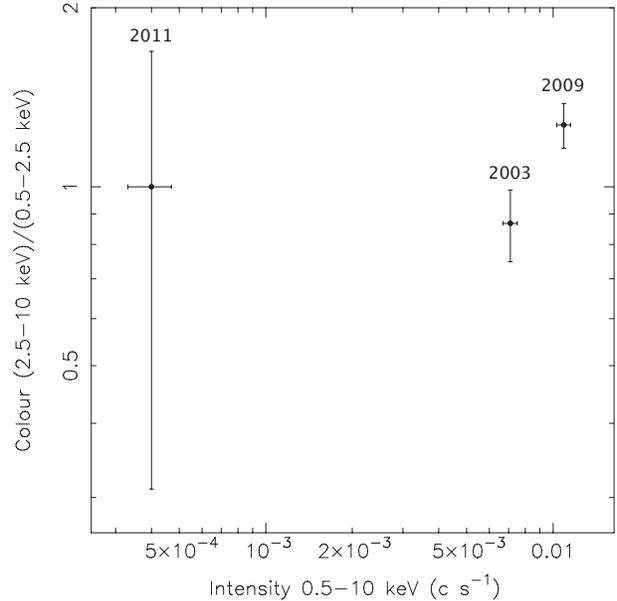}
    \end{center}
\caption[]{{Observed \chan/ACIS-S intensity versus colour of \exo\ during three different quiescent epochs. }}
 \label{fig:colour}
\end{figure}

The spectral analysis is inconclusive as to whether or not the spectral shape changed between the 2003 and 2009 data sets. We therefore carried out a colour-intensity study. For the present purpose we choose two different energy bands of 0.5--2.5 and 2.5--10 keV. We determine the ratio of \chan/ACIS-S counts in the two bands (2.5--10/0.5--2.5 keV) and compare this to the intensity in the full energy band (0.5--10 keV). The results are displayed in Fig.~\ref{fig:colour}. The model-independent colour-intensity study suggests that the X-ray emission was harder and brighter in 2009 when compared to the 2003 data. For completeness we included the 2011 observations (discussed below), but due to the large statistical uncertainties we cannot draw any conclusions about possible changes in the spectral shape compared to the other two epochs (Fig.~\ref{fig:colour}).

\subsection{The 2011 data: flux estimates and constraints}\label{subsec:2011}
Using PWDETECT we obtain a tentative detection with a statistical significance of $3.5\sigma$ at the position of \exo\ for both 2011 observations. In the 2011-February observation there are 11 photons detected within a $1.25''$ circular region centred on the source position. Placing three regions of similar dimensions at source-free locations near the cluster core suggest that $\sim 5$ photons are expected from the background. Applying the prescription for low number statistics of \citet{gehrels1986}, we infer that 12 net photons are detected at 90\% confidence level. This implies a count rate of $\sim 4.0 \times 10^{-4}~\cnts$ (0.5--10 keV). 

In 2011 April there are 13 photons detected at the location of \exo, whereas 8 are collected from our selected background regions. Using \citet{gehrels1986}, we find that a total of 15 net photons are detected at 90\% confidence level, implying a count rate of $\sim 3.8 \times 10^{-4}~\cnts$. It is possible that the excess of photons is due to a statistical fluctuation in the background or a different, overlapping X-ray source. In this case, these count rates can be regarded as upper limits on the intensity of \exo. 

Although the low number of photons does not allow detailed spectral analysis, we can obtain some estimates on the X-ray flux that corresponds to these count rates. We group the combined (background-corrected) 2011 spectrum to a minimum of one photon per bin and fit it in \textsc{XSpec} using W-statistics \citep[an adapted version of Cash's statistics that allows for background subtraction;][]{wachter1979}. 

Assuming that the observed decrease in intensity was completely due to a lower powerlaw normalisation, i.e., adopting the average spectral shape inferred from the 2003 and 2009 data of $N_H=1.2\times10^{22}~\nh$ and $\Gamma=1.5$ (Table~\ref{tab:spec}), we obtain a 0.5--10 keV flux of $F_X\sim1.1 \times 10^{-14}~\flux$ for the 2011 data (cstat=44 for 33 dof). This translates into a luminosity of $L_q \sim 4 \times 10^{31}~\lum$, which would indicate a drop in intensity by a factor of $\sim20-25$ compared to the 2003 and 2009 observations. 

Given the hints of spectral variability between the 2003 and 2009 data sets (Sec.~\ref{subsec:spec}), it may be more likely that the powerlaw index and/or hydrogen column density was different in 2011. For a realistic range in spectral parameters of $N_H=(0.5-2.0)\times10^{22}~\nh$ and $\Gamma=1-3$, we find a 0.5--10 keV flux range of $F_X\sim(0.5-3.4) \times 10^{-14}~\flux$ for the 2011 data, or a luminosity of   
$L_q \sim (2-12) \times 10^{31}~\lum$. We thus conclude that the powerlaw flux varied at least by a factor of $\sim 6$ compared to 2003/2009. If the excess of photons seen during the 2011 observations is not related to the source, however, these estimates represent upper limits and hence the flux variation can be as large as a factor $\gtrsim
50$.

To investigate whether the 2011 data can put further constrains on the thermal quiescent emission, we added an NSATMOS component to the powerlaw fit for $N_H=1.2\times10^{22}~\nh$ and $\Gamma=1.5$. As for the 2003 and 2009 data, we assume $M_{NS}=1.4~\Msun$, $R_{NS}=10$~km, $D=5.5$~kpc and a thermal model normalisation of unity. This way, we obtain an upper limit on the neutron star temperature of $kT^{\infty}\lesssim42$~eV (cstat=44 for 32 dof). The corresponding constraint on the thermal bolometric luminosity, obtained by extrapolating the model fit to the 0.01--100 keV energy range, is $L^{\mathrm{th}}_{\mathrm{q,bol}}\lesssim 7\times 10^{31}~\lum$. The  estimates for the 2011 observations are consistent with the constraints obtained from analysis of the 2003 and 2009 spectral data (Sec.~\ref{subsec:spec}). 

\begin{table*}
\caption{Results from spectral fitting.}
\begin{threeparttable}
\begin{tabular}{c c c c c c c c c}
\hline \hline
Year & $N_H$ & $\Gamma$ & $kT^{\infty}_{\mathrm{eff}}$ & $\chi_{\nu}^2$/cstat (dof) & $F_X$ & $L_q$ & $L^{\mathrm{th}}_{\mathrm{q,bol}}$ & Thermal \\
& ($10^{22}~\nh$) &  & (eV) &  & ($10^{-13}~\flux$) & \multicolumn{2}{c}{($10^{32}~\lum$)} & fraction\\
\hline
\multicolumn{9}{c}{Individual fits: PHABS(POWERLAW)} \\
\hline
2003 & $1.4\pm0.5$ & $1.9\pm0.5$ &  & 0.2 (13) & $2.2^{+0.6}_{-0.3}$ & $8.0^{+2.5}_{-1.1}$ & &  \\
2009 & $1.1\pm0.3$ & $1.3\pm0.3$ & & 1.3 (22) & $2.7\pm0.2$ & $9.8\pm0.7$  &  &  \\
2011 & $1.2$ fix & $1.5$ fix &  & 44 (33) & $0.11\pm0.04$ & $0.4\pm0.2$ & & \\ 
\hline
\multicolumn{9}{c}{Individual fits: PHABS(POWERLAW+NSATMOS)} \\
\hline
2003 & $1.7\pm0.7$ & $1.9\pm0.5$ & $\lesssim89$ & 0.2 (12) & $2.8^{+1.7}_{-1.0}$ & $10^{+4}_{-6}$ & $\lesssim14$ & $\lesssim55\%$ \\
2009 & $1.5\pm0.6$ & $1.3\pm0.3$ & $\lesssim85$ & 1.3 (21) & $3.6\pm1.1$ & $13\pm4$ & $\lesssim12$ & $\lesssim42\%$ \\
2011 & $1.2$ fix & $1.5$ fix & $\lesssim42$ & 44 (32) & $0.11\pm0.05$ & $0.4\pm0.2$ & $\lesssim0.7$ & $\lesssim34\%$ \\ 
\hline
\multicolumn{9}{c}{Simultaneous fit: PHABS(POWERLAW), all parameters tied} \\
\hline
2003 & \multirow{2}{*}{$1.3\pm0.3$} &  \multirow{2}{*}{$1.6\pm0.3$}  & & \multirow{2}{*}{1.3 (38)} & \multirow{2}{*}{$2.3\pm0.2$} & \multirow{2}{*}{$8.3\pm0.7$} & \\
2009 & & & & &  &  & \\
\hline
\multicolumn{9}{c}{Simultaneous fit: PHABS(POWERLAW), $N_H$ and $\Gamma$ tied} \\
\hline
2003 & \multirow{2}{*}{$1.2\pm0.3$} &  \multirow{2}{*}{$1.5\pm0.3$}  & & \multirow{2}{*}{1.0 (37)} & $2.0\pm0.2$ & $7.2\pm0.7$ & \\
2009 & & & & & $2.7\pm0.2$ & $9.8\pm0.7$ & \\
\hline
\multicolumn{9}{c}{Simultaneous fit: PHABS(POWERLAW), $\Gamma$ tied} \\
\hline
2003 & $1.1\pm0.3$ & \multirow{2}{*}{$1.5\pm0.3$} & & \multirow{2}{*}{1.0 (36)}  & $1.9\pm0.2$ & $6.9\pm0.7$ & \\
2009 & $1.3\pm0.3$ &  &  &  & $2.8\pm0.2$ & $10\pm1$ & \\
\hline
\multicolumn{9}{c}{Simultaneous fit: PHABS(POWERLAW), $N_H$ tied} \\
\hline
2003 & \multirow{2}{*}{$1.2\pm0.3$} & $1.7\pm0.3$ & & \multirow{2}{*}{0.9 (36)}  & $2.0\pm0.3$ & $7.2\pm1.1$ & \\
2009 & & $1.4\pm0.3$ &  &  & $2.8\pm0.2$ & $10\pm1$ & \\
\hline
\end{tabular}
\label{tab:spec}
\begin{tablenotes}
\item[]Note. All quoted errors refer to $90\%$ confidence levels. The 2011 data was fitted using W-statistics (an adapted version of Cash's statistics that allows for background subtraction), while we employed $\chi^{2}$-statistics for the 2003 and 2009 data. $F_X$  gives the total unabsorbed model flux in the 0.5--10 keV band and $L_{q}$ represents the corresponding luminosity assuming $D=5.5~$kpc. For the model fits that include a thermal component (NSATMOS) we give the thermal bolometric luminosity ($L^{\mathrm{th}}_{\mathrm{q,bol}}$) and the fractional contribution of the thermal component to the total unabsorbed 0.5--10 keV flux (last column). 
\end{tablenotes}
\end{threeparttable}
\end{table*}

\section{Discussion}\label{sec:discuss}
We use four \chan\ observations of the globular cluster Terzan 5 to study the quiescent spectral and variability properties of the transient neutron star LMXB \exo, which was active in 2000 and 2011. The observations were carried out in 2003, 2009 and 2011. No intervening accretion outbursts have been observed from the cluster during this epoch. 

In 2003, \exo\ is detected at a 0.5--10 keV luminosity of $L_q\sim8\times10^{32}~\lum$, while its intensity was $\sim30\%$ higher 6 years later in 2009 ($L_q\sim1\times10^{33}~\lum$). Apart from flux variations between the two epochs, the source count rate varies up to a factor of $\sim3$ during the individual observations, i.e., on a time scale of hours. On both occasions, the X-ray spectrum is best described by a simple absorbed powerlaw model. The source spectrum is thus dominated by hard, non-thermal emission and lacks the thermal emission component that is seen for many (non-pulsating) neutron star LMXBs at such quiescent luminosities. This implies that all observed variability can be ascribed to the hard spectral component. Our spectral fits suggest that a higher powerlaw normalisation alone can account for the increased flux observed in 2009 compared to 2003. However, a colour-intensity study indicates that the X-ray spectrum likely changed between the two epochs, being spectrally harder (thus possibly more absorbed) in 2009 than in 2003. 

While \exo\ is one of the brightest sources in Terzan 5 during the 2003 and 2009 observations, we find only a small ($3.5\sigma$) excess of photons at the source location during the two observations carried out in 2011. For these tentative detections we estimate a 0.5--10 keV luminosity of $L_q\sim4\times10^{31}~\lum$. If the excess of photons seen during the 2011 observations is not related to the source, this should be considered an upper limit. We conclude that compared to 2003 and 2009, the powerlaw flux had faded at least by a factor of $\sim6$ in 2011, but possibly as much as a factor $\gtrsim 50$. 

By adding a neutron star atmosphere model NSATMOS to the powerlaw spectral fit of the 2011 data, we constrain the neutron star temperature to $kT^{\infty}\lesssim42$~eV and the thermal bolometric luminosity to $L^{\mathrm{th}}_{q,\mathrm{bol}}\lesssim7\times10^{31}~\lum$. These results are consistent with the constraints inferred from the absence of thermal emission during the 2003 and 2009 data sets \citep[Table~\ref{tab:spec}; see also][]{wijnands2005}.

\subsection{Constraints on the thermal emission}\label{sec:thermal}

Whereas neutron stars cool primarily via neutrino emissions from their dense cores and via photon radiation from the stellar surface, episodes of mass-accretion cause heating of the neutron star. Accretion of matter induces a chain of nuclear reactions that deposit heat deep inside the neutron star crust, which is thermally conducted over the entire stellar body. In $\sim10^{4}$~yr, a neutron star reaches a thermal steady state in which heating due to the accretion of matter is balanced by cooling via neutrino emission from the stellar core and photon radiation from the surface \citep[][]{brown1998,colpi2001}. 

According to the minimal cooling paradigm, there is an unescapable rate of neutrino emissions that carries away energy and cools the neutron star \citep[][]{page2004,page2006}. If a neutron star is particularly massive, however, its large central density might open up the threshold for more efficient neutrino emission processes that enhance the cooling rate \citep[][]{yakovlev2004}. Such neutron stars are therefore expected to be relatively cold. In principle, measurements of the thermal surface radiation can constrain the rate of neutrino emissions (i.e., standard/slow versus enhanced) from the neutron star core \citep[][]{yakovlev03,heinke2009}.

For slow neutrino cooling, deep crustal heating should provide an incandescent surface emission that is set by the time-averaged mass-accretion rate of the neutron star as: $L^{\mathrm{th}}_{\mathrm{q,bol}} = \langle \dot{M}_{\mathrm{long}} \rangle Q_{nuc} / m_u$ \citep[e.g.,][]{brown1998,colpi2001}. Here, $Q_{nuc} \sim 2$ MeV is the nuclear energy deposited in the crust per accreted nucleon \citep[e.g.,][]{haensel2008}, $m_u=1.66\times10^{-24}$~g is the atomic mass unit and $\langle \dot{M}_{\mathrm{long}} \rangle$ is the long-term mass-accretion rate of the system averaged over $\sim10^{4}$~yr. If the thermal luminosity is found to be much fainter, this is indicative of a relatively massive neutron star that undergoes enhanced neutrino cooling in its interior. We can estimate the thermal quiescent luminosity expected from deep crustal heating for \exo, and compare this with our observational constraints.

\citet{heinke2003} estimate a bolometric accretion luminosity of $L_{acc}\sim 3 \times 10^{37}~\lum$ by analysing \chan\ and \rxte\ data obtained during the 2000 outburst of \exo. For a bolometric accretion luminosity given by $L_{acc} = (G M_{NS}/R_{NS}) \langle \dot{M}_{ob} \rangle$, we can estimate that the mass-accretion rate during the 2000 outburst was thus $\langle \dot{M}_{ob} \rangle \sim 2 \times 10^{17}~\mathrm{g~s}^{-1}$ or $\sim 3 \times10^{-9}~\mdot$ (assuming a canonical neutron star with $M_{NS} = 1.4~\Msun$ and $R_{NS} = 10$~km). The time-averaged mass-accretion rate of the neutron star ($\langle \dot{M}_{\mathrm{long}} \rangle$) may be approximated by multiplying the average outburst accretion rate ($\langle \dot{M}_{ob} \rangle$) with the duty cycle of the binary, i.e., the ratio of the outburst duration and the recurrence time. The recurrence time of \exo\ is not well constrained, but the historic activity of Terzan 5 provides some rough estimates.

In the past 30 years 8 distinct outbursts have been observed from the cluster, of which two could be ascribed to \exo\ with certainty (2000 and 2011; see Table~\ref{tab:history}). If this is representative for the long-term accretion history of the source, the recurrence time of the system would be $\sim11$~yr. For a typical outburst duration of $\sim2$~months (0.17~yr) the implied duty cycle would be $\sim1\%$. This yields a time-averaged mass-accretion rate of $\langle \dot{M}_{\mathrm{long}} \rangle \sim 2\times 10^{15}~\mathrm{g~s}^{-1}$ ($\sim3\times10^{-11}~\mdot$). For these estimates, deep crustal heating should generate a quiescent bolometric luminosity of $L^{\mathrm{th}}_{\mathrm{q,bol}} \sim 4 \times10^{33}~\lum$. This value would increase if more than two of the Terzan 5 outbursts were due to activity of \exo. 

The above crude estimate suggests that the quiescent thermal luminosity expected to arise from deep crustal heating is within a factor of a few of the upper limits inferred from the 2003 and 2009 \chan\ data sets ($L^{\mathrm{th}}_{\mathrm{q,bol}} \lesssim 1\times10^{33}~\lum$; Sec.~\ref{subsec:spec}), but a factor $\gtrsim60$ higher than the constraints we obtain from analysis of 2011 quiescent data ($L^{\mathrm{th}}_{\mathrm{q,bol}} \lesssim 7\times10^{31}~\lum$; Sec.~\ref{subsec:2011}). If the low luminosity observed in 2011 is intrinsic to the source and not caused by temporary obscuration of the X-ray emission (see Sec.~\ref{subsec:var}), the neutron star is thus colder than expected for a duty cycle of 1\%. This might imply that the source on average has a longer recurrence time and that the behavior seen in the past 30 years is not representative for its long-term activity. When slow neutrino emission processes are operating in the core, a thermal bolometric quiescent luminosity of $L^{\mathrm{th}}_{\mathrm{q,bol}} \lesssim 7\times10^{31}~\lum$ would require a time-average mass-accretion rate of $\langle \dot{M}_{\mathrm{long}} \rangle \lesssim 4\times 10^{13}~\mathrm{g~s}^{-1}$ ($\lesssim 6\times10^{-13}~\mdot$). If the observed 2000 outburst is typical for the long-term accretion history of \exo, its duty cycle would need to be $\lesssim 0.02\%$, meaning that the source should reside $\gtrsim 900$~yr in quiescence in between subsequent outbursts. 

Alternatively, the neutron star in \exo\ might undergo enhanced core cooling, as was previously proposed by \citet{wijnands2005}. The same conclusion was drawn for the other transient neutron star LMXB in Terzan 5, \igr, based on its quiescent thermal luminosity \citep[][]{deeg_wijn2011}. This would suggest that at least one of the two neutron star LMXBs undergoes enhanced core cooling or that both have a long recurrence time. The latter scenario would imply that other historic outbursts of Terzan 5 were likely caused by one or more additional transient sources.

\subsection{Quiescent variability}\label{subsec:var}
Investigation of the four observations discussed in this work reveals that the quiescent emission of \exo\ varies both on short and long timescales. Summarizing: during the 2003 and 2009 observations the source intensity changed by a factor of a few on a timescale of hours. The X-ray flux varied by $\sim30\%$ between 2003 and 2009, i.e., over a timescale of 6 years, whereas the powerlaw flux had decreased by a factor of $\gtrsim 6$ (and depending on the spectral parameters possibly as much as a factor $\gtrsim 50$) in the 19 months separating the 2009 and first 2011 observation. There is no evidence for variability between the two 2011 observations, which were separated by $\sim2$~months. It is of note that the source entered a new accretion outburst five months later in 2011 October \citep[][]{altamirano2011a}. Several (weakly magnetised) transient neutron star LMXBs have been found to exhibit changes in their quiescent luminosity and this variability has been interpreted in different ways. 

For a group of five sources, \ks\ \citep{wijnands2002,cackett2010}, \mxb\ \citep{wijnands2004,cackett2008}, \xte\ \citep{fridriksson2010,fridriksson2011}, \qpexo\ \citep{degenaar09_exo1,degenaar2010_exo2,diaztrigo2011} and \igr\ \citep[the other LMXB in Terzan 5;][]{degenaar2011_terzan5_3} a gradual decrease in thermal X-ray emission, extending to many years, is observed after the end of an accretion outburst. This variability has been ascribed to cooling of the neutron star crust that was heated due to the accretion of matter \citep[e.g.,][]{rutledge2002}. For one of these sources, \xte, episodes of increased thermal emission were found superimposed on the decaying trend. These involved a correlated increase in the contribution of the powerlaw spectral component and have therefore been attributed to temporal low-level accretion onto the neutron star surface \citep{fridriksson2010,fridriksson2011}. Neutron star crust cooling cannot be invoked as an explanation for the quiescent variability of \exo, since its quiescent emission spectrum is fully dominated by hard non-thermal emission.

Both Aql X-1 and Cen X-4 have been extensively studied in quiescence, using observations spanning decades, which has revealed intensity variations occurring on time scales of minutes to years \citep[][]{rutledge1999,rutledge2002_aqlX1,campana2003_aqlx1,campana2004,cackett2010_cenx4,cackett2011}. For Cen X-4, a recent study showed that both spectral components need to be variable to explain the observed intensity variations \citep[][]{cackett2010_cenx4}. In particular, it was found that the ratio of the powerlaw to thermal flux remained approximately constant, while the total flux changed by a factor of $\gtrsim4$. This suggests a direct connection between the two spectral components and led to the interpretation of residual accretion acting in quiescence \citep[][]{cackett2010_cenx4}. 
Studies of Aql X-1 were not conclusive about which of the spectral components was underlying the observed variability. Both accretion onto the neutron star surface with a variable mass-inflow rate, as well as the variable interaction between the pulsar relativistic wind and matter outflowing from the companion star were suggested to explain the observed variability \citep[][]{rutledge1999,rutledge2001,rutledge2002_aqlX1,campana2003_aqlx1,cackett2011}.

The accreting millisecond X-ray pulsars NGC 6440 X-1 and \sax\ both displayed quiescent flux variations by a factor $\lesssim2$ that could be entirely attributed to the powerlaw spectral component \citep{cackett2005,campana2008_saxj1808,heinke2009}. The behaviour of \exo\ is thus perhaps most reminiscent of these two neutron stars, although its variation in powerlaw flux appears to be stronger (up to a factor $\gtrsim 6$). It is of note that some transient LMXBs harbouring a black hole primary exhibit X-ray intensity variations in their quiescent state that are of similar magnitude as observed for \exo\ \citep[][]{kong2002,hynes2004}. These systems display pure powerlaw spectra in quiescence, without any thermal emission component. The mechanism underlying the variations in powerlaw flux seen in black hole LMXBs might also produce the observed variations in the neutron star systems. This would necessary imply, however, that the powerlaw emission cannot be related to any distinctive properties of a neutron star, such as a solid surface and a magnetic field.

Accretion is known to cause considerable variability on long timescales, driving the outburst and quiescence cycles of transient LMXBs, as well as on short time scales during outburst episodes. Continued low-level accretion therefore seems an attractive explanation for the observed variability of \exo. The source might have been accreting at 0.5--10 keV luminosities of $L_q\sim(8-10)\times10^{32}~\lum$ when it was observed in 2003 and 2009, while its true quiescent luminosity (i.e., the emission level when all accretion activity has ceased) is lower and possibly represented by the 2011 observations ($L_q\sim4\times10^{31}~\lum$, 0.5--10 keV). However, it may not be straightforward to explain the large and relatively rapid drop in X-ray intensity that occurred between 2009 and 2011 within this framework. This would require a strong reduction of the accretion flow in the 19 months separating the two observations, while it had restored again only $\sim5$~months later when the source exhibited a new accretion outburst \citep[2011 late-October;][]{altamirano2011a}.

Possibly, the non-thermal X-ray emission was not intrinsically reduced but rather obscured during the 2011 observations. \exo\ displayed prominent dipping activity during the first $\sim50$ days of its 2000 outburst \citep[][]{markwardt2000_2}. During the dips the X-ray intensity was reduced by a factor of a few, indicating that the central X-ray source was temporarily obscured, although the dips disappeared and did not return in the second half of the outburst (Altamirano et al. in preparation). The same process may cause obscuration of the X-ray emission in quiescence. For example, the occurrence of a new outburst of \exo\ in late-2011 indicates that there was likely a (cold) accretion disk present during the 2011 quiescence observations (which were separated by $\sim75$~days; see Table~\ref{tab:obs}) that may have partly blocked the central X-ray source and hence caused the faded X-ray emission compared to the 2003 and 2009 data.

Our colour-intensity study and spectral analysis suggests that the hydrogen column density may have varied between the 2003 and 2009 observations, which could lend support to the presence of an obscuring medium in quiescence. This interpretation would require the powerlaw emission to be produced close to the central X-ray source. An origin at the shock front between the pulsar wind and matter flowing out from the companion star would then be less likely than accretion of matter onto the neutron star's magnetic field or surface. We note that depending on the exact spectral parameters, the decrease in quiescent powerlaw flux might be as large as a factor of $\gtrsim50$ (Sec.~\ref{subsec:2011}), which would be much stronger than the intensity variations seen during the dipping activity in the 2000 outburst of \exo.

\exo\ adds to the (growing) list of neutron star LMXBs that show considerable X-ray variability during quiescence. New \chan\ observations of Terzan 5 have been planned for the coming year. These allow to further study the quiescent properties of \exo\ and can potentially shed more light on the nature of its strong quiescent X-ray variability.

~\\
\noindent {\bf Acknowledgements.}\\
This work was supported by the Netherlands Research School for Astronomy (NOVA). ND is supported by NASA through Hubble Postdoctoral Fellowship grant number HST-HF-51287.01-A from the Space Telescope Science Institute, which is operated by the Association of Universities for Research in Astronomy, Incorporated, under NASA contract NAS5-26555. RW acknowledges support from a European Research Council (ERC) starting grant. This work made use of the \chan\ public data archive, \maxi\ data provided by RIKEN, JAXA and MAXI team, as well as results provided by the ASM/\rxte\ teams at MIT and at the \rxte\ SOF and GOF at NASA's GSFC. The authors are grateful to the anonymous referee for thoughtful comments that helped improve this manuscript.

\label{lastpage}
\end{document}